\pdfoutput=1
\documentclass[a4paper,11pt]{article}

\usepackage[
    margin=0.8in
    ]{geometry}

\usepackage{hyperref}
\hypersetup{
    colorlinks=true,
    linktocpage=true,
    linkcolor=blue,
    urlcolor=blue,
    citecolor=blue,
    filecolor=blue,
    bookmarksopen=true,
    }

\usepackage{amsmath,amssymb,mathtools,mathrsfs}
\usepackage[italicdiff]{physics}
\mathtoolsset{showonlyrefs,showmanualtags}
\numberwithin{equation}{section}
\allowdisplaybreaks[4]

\usepackage[
    backend=biber,
    bibstyle=phys,
    sorting=none,
    biblabel=brackets,
    citestyle=numeric-comp,
    pageranges=false,
    eprint=true,
    url=true,
    ]{biblatex}
\AtEveryBibitem{
    \clearfield{note}
    \clearname{editor}
}
\DeclareFieldFormat{url}{[\href{#1}{iNSPIRE}]}
\DeclareSourcemap{
    \maps[datatype=bibtex, overwrite=true]{
        \map{
            \step[fieldsource=journal,
                match=\regexp{Advances\sin\sTheoretical\sand\sMathematical\sPhysics},
                replace={Adv.Theor.Math.Phys.}]
            \step[fieldsource=journal,
                match=\regexp{Journal\sof\sHigh\sEnergy\sPhysics},
                replace={JHEP}]
            \step[fieldsource=journal,
                match=\regexp{Nuclear\sPhysics\sB},
                replace={Nucl.Phys.B}]
            \step[fieldsource=journal,
                match=\regexp{Physics\sLetters\sB},
                replace={Phys.Lett.B}]
            \step[fieldsource=journal,
                match=\regexp{Physical\sReview\sD},
                replace={Phys.Rev.D}]
            \step[fieldsource=journal,
                match=\regexp{Progress\sof\sTheoretical\sPhysics},
                replace={Prog.Theor.Phys.}]
            \step[fieldsource=journal,
                match=\regexp{Progress\sof\sTheoretical\sand\sExperimental\sPhysics},
                replace={PTEP}]
            \step[fieldsource=journal,
                match=\regexp{Theoretical\sand\sMathematical\sPhysics},
                replace={Theor.Math.Phys.}]
            }
        }
    }
\bibliography{reference.bib}

\makeatletter
\renewcommand\thefootnote{\@fnsymbol\c@footnote}%
\renewcommand{\maketitle}{%
    \newpage
    \begin{flushright}
        UTHEP- 777
    \end{flushright}
    \null
    \vskip 2em%
    \begin{center}%
    \let \footnote \thanks
    {\LARGE\textbf{\@title}\par}%
    \vskip 1.5em%
    {\large
        \lineskip .5em%
        \begin{tabular}[t]{c}%
        \@author
    \end{tabular}\par}%
    \vskip 1em%
    {\large \@date}%
    \end{center}%
    \par
    \@thanks
    \vskip 1.5em}
\makeatother
\usepackage{authblk}

\title{The classical solutions with $k_-=0$ in Kaku theory}
\author{Yuji Ando\thanks{E-mail: \href{mailto:ando@het.ph.tsukuba.ac.jp}{ando@het.ph.tsukuba.ac.jp}}}
\affil{
Degree Programs in Pure and Applied Sciences,\\
Graduate School of Science and Technology, University of Tsukuba,\\
Tsukuba, Ibaraki 305-8571, Japan}
\date{}

\usepackage{tikz}
\usetikzlibrary{intersections,calc}

\begin{document}
\maketitle
\renewcommand\thefootnote{\arabic{footnote}}
\setcounter{footnote}{0}
\begin{abstract}
    We consider Kaku theory as introduced in M. Kaku, \href{https://doi.org/10.1016/0370-2693(88)91102-1}{Phys. Lett. B 200, 22 (1988)} and investigate classical solutions. In particular, we obtain that the equation of motion with the restriction $k_-=0$ in the Kaku theory is equivalent to the equation of motion in Witten theory. Because of this property, some solutions including the tachyon vacuum solution in the Witten theory satisfy also the equation of motion in the Kaku theory. In addition, we confirm that the cohomology around the tachyon vacuum solution is trivial also in the Kaku theory.
\end{abstract}
\thispagestyle{empty}
\newpage
\setcounter{page}{1}
\tableofcontents
\section{Introduction and Summary}
String field theory is a candidate for a non-perturbative formulation of string theory, and it was expected that string field theory can potentially be used for proving Sen's conjecture \cite{Sen:1999xm} and studying non-perturbative effects of string theory and so on. In particular, if we construct the tachyon vacuum solution of open string field theory and study the behavior of open and closed strings around the solution, we understand the behavior of strings on the open string background that is not able to be perturbatively considered.

One of covariant open bosonic string field theories is Witten theory and the action can be written as
\begin{align}
    S_\mathrm{W}=\frac{1}{2}\omega(\Psi,Q\Psi)+\frac{1}{3}\omega(\Psi,m_2^\mathrm{W}(\Psi,\Psi)),
\end{align}
and $m_2^\mathrm{W}$ is the string product which defines the Witten vertex \cite{Witten:1985cc}. In this theory, many classical solutions including the tachyon vacuum solution were constructed, and using the tachyon vacuum solution, we are able to confirm Sen's conjecture as we expect \cite{Schnabl:2005gv,Okawa:2006vm,Erler:2009uj,Erler:2010zza,Ellwood:2006ba}. On the other hand, it is unclear how closed strings appear around the tachyon vacuum solution.

One of the other theories is the theory using the string products $m_2^\mathrm{lc}$ and $m_3^\mathrm{lc}$ which define the cubic and quartic light-cone vertices. The light-cone vertex takes momentum $k_-$ as the length parameter of string fields and the action can be written as 
\begin{align}
    S=\frac{1}{2}\omega(\Psi,Q\Psi)+\frac{1}{3}\omega(\Psi,m_2^\mathrm{lc}(\Psi,\Psi))+\frac{1}{4}\omega(\Psi,m_3^\mathrm{lc}(\Psi,\Psi,\Psi))
\end{align}
which is called $\alpha=p^+$ HIKKO theory or Kugo-Zweibach theory \cite{Hata:1986jd,Hata:1986ke,Kugo:1992md}. Through the action, the dynamics of closed strings are not described. To describe them, other actions are used \cite{Saitoh:1988ew,Kugo:1997rm,Asakawa:1998em}. Thus given a tachyon vacuum solution, it is clear how closed strings appear around it in this theory. In contrast with the Witten theory, classical solutions have not yet been found in this theory, and we are not able to confirm Sen's conjecture. One of the reasons is the length parameter. Many of the solutions that we are interested in contain states with $k_-=0$. However, we face a problem when we consider states with $k_-=0$ in this theory, and we cannot construct these solutions unless the problem is solved.

Our goal is to find the theory which can be used to construct classical solutions and understand how closed strings appear around the tachyon vacuum solution. We focus on the Kaku theory as a candidate for such theory. The action can be written as
\begin{align}
    S^l=\frac{1}{2}\omega(\Psi,Q\Psi)+\frac{1}{3}\omega(\Psi,m_2^l(\Psi,\Psi))+\frac{1}{4}\omega(\Psi,m_3^l(\Psi,\Psi,\Psi)),
\end{align}
and $m_2^l$ and $m_3^l$ are string products which define the cubic and quartic Kaku vertex introduced in \cite{Kaku:1987jx}. The Kaku vertex has the Chan-Paton parameter $l$ in addition to the length parameter $k_-$, and the specific choices $l\to\infty$ and $l\to0$ correspond to the Witten vertex and the light-cone vertex, respectively. Thus this theory is intermediate between them, and it may be possible in this theory to construct classical solutions and understand how closed strings appear.

In this paper, we explore the tachyon vacuum solution in the Kaku theory. To facilitate our exploration, we examine how the state with $k_-=0$ is treated. As we expect, owing to the Chan-Paton parameter, we can consider the states with $k_-=0$ without any problem. Luckily, we obtain that the equation of motion with the restriction $k_-=0$ in the Kaku theory is equivalent to the equation of motion in the Witten theory. Unfortunately, we have not yet demonstrated that the physical interpretation of the solutions is the same. For example, there is a possibility that some solutions which are not gauge equivalent in the Witten theory are gauge equivalent in the Kaku theory. However in this paper, we show that, at least, the energy of the solution has the same value in either the Witten theory or the Kaku theory. Moreover, we confirm the existence of the homotopy operator for the tachyon vacuum solution in the Kaku theory.

This paper is organized as follows. In section \ref{light-cone}, we review the light-cone vertex and the problem for states with $k_-=0$. In section \ref{Kaku}, we introduce the Kaku vertex and confirm that we do not face the problem in the Kaku theory when we consider states with $k_-=0$. In section \ref{cla}, we investigate classical solutions in the Kaku theory. Finally, in section \ref{summary}, we present the summary.
\section{Light-cone vertex\label{light-cone}}
In this section, we review the light-cone vertex and the problems we face for states with $k_-=0$. See appendix B of \cite{Erler:2020beb} for more details.

Given an open string state $\ket{A}$, we can define a corresponding boundary operator $A(0)$ such that
\begin{align}
    \ket{A}=A(0)\ket{0}
\end{align}
where $\ket{0}$ is the $SL(2,\mathbb{R})$ invariant vacuum. Let us write the vertex as a correlation function
\begin{align}
    V_3^\mathrm{lc}(A,B,C)&=\ev{f_{(3,1)}^\mathrm{lc}\circ A(0)f_{(3,2)}^\mathrm{lc}\circ B(0)f_{(3,3)}^\mathrm{lc}\circ C(0)}_\mathrm{UHP}\\
    &=(-1)^{\abs{A}+\abs{B}}\omega(A,m_2^\mathrm{lc}(B,C))
\end{align}
where $\abs{A}$ and $\abs{B}$ are defined to be their Grassmann parity plus 1\footnote{We follow the notation in \cite{Erler:2015uba}}. The maps $f_{(3,r)}^\mathrm{lc}$ are derived from Mandelstam mapping \cite{Mandelstam:1973jk,LeClair:1988sp,LeClair:1988sj}. Suppose that the states $A,B$ and $C$ have momenta $k_-^A,k_-^B$ and $k_-^C$ with $k_-^A,k_-^B>0,k_-^C<0$ respectively. This can be done without loss of generality. The momenta satisfy
\begin{align}
    \abs{k_-^A}+\abs{k_-^B}=\abs{k_-^C}\label{conservation}
\end{align}
as a result of momentum conservation. We assign momentum $k_-$ as the length of the open string propagator strip (Figure \ref{light-conestring}) and the cubic light-cone vertex is represented as Figure \ref{cubiclight-cone}. We call the surface in Figure \ref{cubiclight-cone} $\Sigma$. Then the Mandelstam mapping from the complex coordinate $u$ on upper half plane(UHP) to the complex coordinate $\rho$ on $\Sigma$ which is normalized by $\abs{k_-^C}$ is given by
\begin{align}
    \rho(u)&=\frac{1}{\pi}\int\dd{u}\frac{u-u_*}{u(u-1)}\label{def:Mandelstam}\\
    &=\frac{\tilde{\alpha}_A}{\pi}\ln(u-1)+\frac{\tilde{\alpha}_B}{\pi}\ln u
\end{align}
where $\tilde{\alpha}_A$ and $\tilde{\alpha}_B$ are defined by
\begin{align}
    \tilde{\alpha}_A\coloneqq\frac{\abs{k_-^A}}{\abs{k_-^C}}\qc\tilde{\alpha}_B\coloneqq\frac{\abs{k_-^B}}{\abs{k_-^C}}\label{def:tildealpha}
\end{align}
and the operators $A,B$ and $C$ are placed respectively at $1,0$ and $\infty$. $u_*=\tilde{\alpha}_B$ is the interaction point which is placed on the boundary.

By appropriate scaling, translation, and exponentiation on $\rho$, we can find the inverse local coordinate maps given by
\begin{align}
    (f_{(3,1)}^\mathrm{lc})^{-1}(u)=\frac{u-1}{\tilde{\alpha}_A}\qty(\frac{u}{\tilde{\alpha}_B})^{\frac{\tilde{\alpha}_B}{\tilde{\alpha}_A}}\qc(f_{(3,2)}^\mathrm{lc})^{-1}(u)=\qty(\frac{1-u}{\tilde{\alpha}_A})^{\frac{\tilde{\alpha}_A}{\tilde{\alpha}_B}}\frac{u}{\tilde{\alpha}_B}\qc(I\circ f_{(3,3)}^\mathrm{lc})^{-1}(u)=u\qty(\frac{\tilde{\alpha}_A}{u+1})^{\tilde{\alpha}_A}\tilde{\alpha}_B^{\tilde{\alpha}_B}
\end{align}
and their derivatives are
\begin{align}
    \eval{\dv{(f_{(3,1)}^\mathrm{lc})^{-1}}{u}}_{u=1}=\tilde{\alpha}_A^{-1}\tilde{\alpha}_B^{-\frac{\tilde{\alpha}_B}{\tilde{\alpha}_A}}\qc\eval{\dv{(f_{(3,2)}^\mathrm{lc})^{-1}}{u}}_{u=0}=\tilde{\alpha}_A^{-\frac{\tilde{\alpha}_A}{\tilde{\alpha}_B}}\tilde{\alpha}_B^{-1}\qc\eval{\dv{(I\circ f_{(3,3)}^\mathrm{lc})^{-1}}{u}}_{u=0}=\tilde{\alpha}_A^{\tilde{\alpha}_A}\tilde{\alpha}_B^{\tilde{\alpha}_B}.\label{derivative:Mandelstam}
\end{align}

Similarly, the quartic light-cone vertex for the case $\mathrm{sgn}(k_-^A,k_-^B,k_-^C,k_-^D)=(+,-,+,-)$ or $(-,+,-,+)$ is defined by
\begin{align}
    V_4^\mathrm{lc}(A,B,C,D)&=\int_0^1\dd{m}\ev{b(v)f_{(4,1)}^\mathrm{lc}\circ A(0)f_{(4,2)}^\mathrm{lc}\circ B(0)f_{(4,3)}^\mathrm{lc}\circ C(0)f_{(4,4)}^\mathrm{lc}\circ D(0)}_\mathrm{UHP}\\
    &=(-1)^{1+\abs{C}}\omega(A,m_3^\mathrm{lc}(B,C,D))
\end{align}
where $b(v)$ is the appropriate $b$-ghost insertion for the measure on moduli space. The maps $f_{(4,r)}^\mathrm{lc}$ are also derived from Mandelstam mapping
\begin{align}
    \rho(u)=\int\dd{u}\frac{u-u_*}{u(u-m)(u-1)}
\end{align}
in the same way as in the cubic light-cone vertex, where the operators $A,B,C$ and $D$ are placed respectively at $1,m,0$ and $\infty$. For the other cases about $k_-$, we define it as
\begin{align}
    V_4^\mathrm{lc}(A,B,C,D)=0.
\end{align}
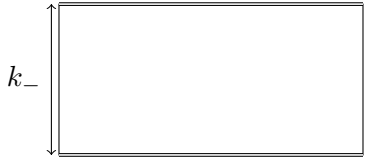
\begin{figure}
    \centering
    \begin{tikzpicture}
        \draw (-2,0)--(2,0)--(2,2)--(-2,2)--cycle;
        \draw[double] (-2,0)--(2,0);
        \draw[double] (2,2)--(-2,2);
        \draw[<->,shift={(-0.1,0)}] (-2,0)--(-2,2) node[midway,left] {$k_-$};
    \end{tikzpicture}
    \caption{An open string propagator with the length $k_-$}\label{light-conestring}
\end{figure}
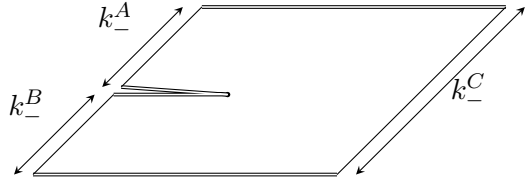
\begin{figure}
    \centering
    \begin{tikzpicture}
        \coordinate (o) at (0,0);
        \coordinate (i) at ($(o)+(pi/4 r:1.5)+(3/2,0)$);
        \fill (i) circle (1pt);
        \coordinate (v) at ($(i)$);
        \coordinate (s1) at ($(o)+(0,0)$);
        \coordinate (p11) at ($(s1)+(0,0)$);
        \coordinate (p12) at ($(p11)+(pi/4 r:1.5)$);
        \coordinate (s2) at ($(p12)+(0,0)$);
        \draw[<->,>=stealth] ($(p11)+(-1/4,0)$)--($(p12)+(-1/4,0)$) node[midway,above left]{$k_-^B$};
        \coordinate (t1) at ($(s2)+(0.1,0.1)$);
        \coordinate (p21) at ($(t1)+(0,0)$);
        \coordinate (p22) at ($(p21)+(pi/4 r:1.5)$);
        \coordinate (t2) at ($(p22)+(0,0)$);
        \draw[<->,>=stealth] ($(p21)+(-1/4,0)$)--($(p22)+(-1/4,0)$) node[midway,above left]{$k_-^A$};
        \coordinate (u1) at ($(t2)+(4,0)$);
        \coordinate (p31) at ($(u1)+(0,0)$);
        \coordinate (p32) at ($(s1)-(t2)+(u1)$);
        \coordinate (u2) at ($(p32)+(0,0)$);
        \draw[<->,>=stealth] ($(p31)+(1/4,0)$)--($(p32)+(1/4,0)$) node[midway,right]{$k_-^C$};
        \draw[double] (s2)--(v)--(t1);
        \draw[double] (t2)--(u1);
        \draw[double] (u2)--(s1);
        \draw (s1)--(s2);
        \draw (t1)--(t2);
        \draw (u1)--(u2);
    \end{tikzpicture}
    \caption{The cubic light-cone vertex when $k_-^A,k_-^B>0,k_-^C<0$.}\label{cubiclight-cone}
\end{figure}

Due to the above definition, the action using the light-cone vertices
\begin{align}
    S=\frac{1}{2}\omega(\Psi,Q\Psi)+\frac{1}{3}\omega(\Psi,m_2^\mathrm{lc}(\Psi,\Psi))+\frac{1}{4}\omega(\Psi,m_3^\mathrm{lc}(\Psi,\Psi,\Psi))
\end{align}
has the $A_\infty$ structure and satisfies the Batalin-Vilkovisky(BV) master equation without higher order vertices e.g. quintic light-cone vertex\footnote{See \cite{Erler:2013xta,Erler:2015uba} for details about $A_\infty$ structure.}.

When we consider the state with $k_-=0$ in this theory, we must be careful. As a concrete example, let us examine the case of $k_-^A=0$ when $A,B$ and $C$ are primary states of weight $h_A,h_B$ and $h_C$, respectively. This corresponds to the limit $\tilde{\alpha}_A\to0$ and $\tilde{\alpha}_B\to1$ from \eqref{conservation} and \eqref{def:tildealpha}. Thus we obtain
\begin{align}
    &\lim_{k_-^A\to0}V_3^\mathrm{lc}(A,B,C)\\
    &=\lim_{\tilde{\alpha}_A\to0}\qty[\tilde{\alpha}_A^{-1}(1-\tilde{\alpha}_A)^{-\frac{1-\tilde{\alpha}_A}{\tilde{\alpha}_A}}]^{-h_A}\qty[\tilde{\alpha}_A^{-\frac{\tilde{\alpha}_A}{1-\tilde{\alpha}_A}}(1-\tilde{\alpha}_A)^{-1}]^{-h_B}\qty[\tilde{\alpha}_A^{\tilde{\alpha}_A}(1-\tilde{\alpha}_A)^{1-\tilde{\alpha}_A}]^{-h_C}\ev{A(1)B(0)C(\infty)}_\mathrm{UHP}.
\end{align}
In the limit, the conformal factor becomes
\begin{align}
    \lim_{\tilde{\alpha}_A\to0}\qty[\tilde{\alpha}_A^{-1}(1-\tilde{\alpha}_A)^{-\frac{1-\tilde{\alpha}_A}{\tilde{\alpha}_A}}]^{-h_A}\qty[\tilde{\alpha}_A^{-\frac{\tilde{\alpha}_A}{1-\tilde{\alpha}_A}}(1-\tilde{\alpha}_A)^{-1}]^{-h_B}\qty[\tilde{\alpha}_A^{\tilde{\alpha}_A}(1-\tilde{\alpha}_A)^{1-\tilde{\alpha}_A}]^{-h_C}=\begin{cases}
        \infty&\qfor h_A<0.\\
        1&\qfor h_A=0.\\
        0&\qfor h_A>0.
    \end{cases}
\end{align}
Therefore for the state $k_-^A=0$, the cubic light-cone vertex is well-defined only if $h_A\ge0$. Moreover, if we consider the case of $k_-^A=k_-^B=k_-^C=0$, the result is not unique depending on how to take the limit. Thus, we cannot treat the states with $k_-=0$ and it is difficult to construct the classical solutions with $k_-=0$ in this theory.
\section{Kaku vertex\label{Kaku}}
As we saw in the previous section, the light-cone vertex is not well-defined for the states with $k_-=0$. To avoid this problem, we consider the theory using the Kaku vertex \cite{Kaku:1987jx}. In this section, we review the Kaku vertex and the Kaku theory. For more details, reader can refer to \cite{Erler:2020beb}.

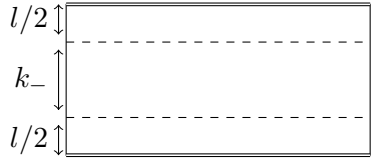
\begin{figure}
    \centering
    \begin{tikzpicture}
        \draw (-2,0)--(2,0)--(2,2)--(-2,2)--cycle;
        \draw[double] (-2,0)--(2,0);
        \draw[double] (2,2)--(-2,2);
        \draw[dashed] (-2,1/2)--(2,1/2);
        \draw[dashed] (-2,3/2)--(2,3/2);
        \draw[<->,shift={(-0.1,0)}] (-2,3/2+0.1)--(-2,2) node[midway,left] {$l/2$};
        \draw[<->,shift={(-0.1,0)}] (-2,1/2+0.1)--(-2,3/2-0.1) node[midway,left] {$k_-$};
        \draw[<->,shift={(-0.1,0)}] (-2,0)--(-2,1/2-0.1) node[midway,left] {$l/2$};
    \end{tikzpicture}
    \caption{An open string propagator with the length $k_-+l$}\label{Kakustring}
\end{figure}

\begin{figure}
    \centering
    \begin{tikzpicture}[scale=1.2]
        \coordinate (o) at (0,0);
        \coordinate (i) at ($(o)+(pi/4 r:1)+(3/2,0.1)$);
        \fill (i) circle (1pt);
        \coordinate (v) at ($(i)+(pi/2 r:1)$);
        \coordinate (s1) at ($(o)+(0,0)$);                
        \coordinate (p11) at ($(s1)+(pi/4 r:{1/3})$);
        \coordinate (p12) at ($(p11)+(pi/4 r:{1*2/3})$);
        \coordinate (s2) at ($(p12)+(pi/2 r:{1})$);
        \draw[<->] ($(s1)+(-1/4,0)$)--($(p11)+(-1/4-0.05,0-0.05)$) node[midway,above left] {$l/2$};
        \draw[<->] ($(p11)+(-1/4+0.05,0+0.05)$)--($(p12)+(-1/4-0.05,0-0.05)$) node[midway,above left] {$k_-^A$};
        \draw[<->] ($(p12)+(-1/4+0.05,0+0.05)$)--($(s2)+(-1/4+0.05,0-0.05)$) node[midway,above left] {$l/2$};
        \coordinate (t1) at ($(s2)+(pi/4 r:1)$);
        \coordinate (p21) at ($(t1)+(-pi/2 r:{1})$);
        \coordinate (p22) at ($(p21)+(pi/4 r:{1*2/3})$);
        \coordinate (t2) at ($(p22)+(pi/4 r:{1/3})$);
        \coordinate (u1) at ($(t2)+(4,0)$);
        \coordinate (p31) at ($(u1)+(-pi*3/4 r:{1/3})$);
        \coordinate (p32) at ($(p31)+(-pi*3/4 r:{1*7/3})$);
        \coordinate (u2) at ($(s1)-(t2)+(u1)$);
        \draw[<->] ($(u1)+(1/4,0)$)--($(p31)+(1/4+0.05,+0.05)$) node[midway,below right] {$l/2$};
        \draw[<->] ($(p31)+(1/4-0.05,-0.05)$)--($(p32)+(1/4+0.05,+0.05)$) node[midway,right] {$k_-^C$};
        \draw[<->] ($(p32)+(1/4-0.05,-0.05)$)--($(u2)+(1/4,0)$) node[midway,below right] {$l/2$};
        \draw[name path=b,double] (s2)--(v)--(t1);
        \draw[double] (t2)--(u1);
        \draw[double] (u2)--(s1);
        \draw (s1)--(p12)--(s2);
        \draw[name path=string2,dotted] (t1)--(p21)--(t2);
        \path[name intersections={of= b and string2}];
        \draw (t1)--(intersection-1);
        \draw (intersection-4)--(t2);
        \draw (u1)--(u2);
        \draw[dash dot] (p12)--(i)--(p21);
        \draw[dashed] (v)--(i);
        \draw[dashed] (p11)--(p32);
        \draw[dashed] (p22)--(p31);
    \end{tikzpicture}
    \caption{The cubic Kaku vertex when $k_-^A,k_-^B>0,k_-^C<0$.}\label{cubicKaku}
\end{figure}
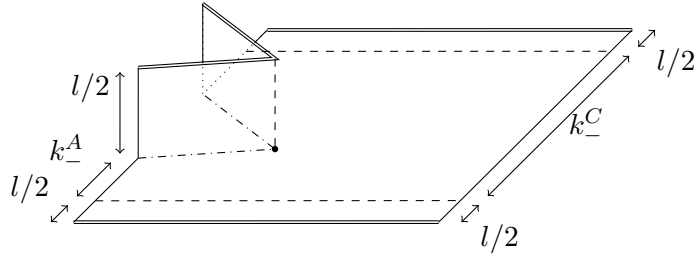

To define the Kaku vertex, we assign $k_-+l$ as the length of the open string propagator using the momentum $k_-$ and the Chan-Paton parameter $l$ (Figure \ref{Kakustring}).
The momentum $k_-$ part of the world-sheet splits and joins in the same way as in the light-cone vertex, and the Chan-Paton parameter $l$ part of it is glued together in the same way as in the Witten vertex. As a result, the cubic Kaku vertex is represented by Figure \ref{cubicKaku}.
We obtain the light-cone vertex in the limit $l\to0$, whereas we obtain the Witten vertex in the limit $l\to\infty$. Thus, the Kaku vertex is intermediate between them.

In this case, let us write the vertex as a correlation function
\begin{align}
    V_3^l(A,B,C)&=\ev{f_{(3,1)}^\mathrm{lc}\circ A(0)f_{(3,2)}^l\circ B(0)f_{(3,3)}^l\circ C(0)}_\mathrm{UHP}\\
    &=(-1)^{\abs{A}+\abs{B}}\omega(A,m_2^l(B.C))
\end{align}
and construct the maps $f_{(3,r)}^l$. These maps are derived from the map
\begin{align}
    \rho(u)=\frac{1}{\pi}\int\dd{u}\frac{\sqrt{(u-u_*)(u-\bar{u}_*)}}{u(u-1)}.\label{def:Kakumap}
\end{align}
As in the previous section, the operators $A,B$ and $C$ are placed respectively at $1,0$ and $\infty$. However, the interaction point $u_*$ can be placed not only on the boundary but also in the bulk.

When we consider \eqref{def:Kakumap}, it is necessary to distinguish whether the interaction point is placed on the boundary or in the bulk. In fact, for $\Im u_*=0$, equation \eqref{def:Kakumap} coincides with \eqref{def:Mandelstam}. On the other hand, assuming $\Im u_*\neq0$, equation \eqref{def:Kakumap} gives
\begin{align}
    \rho(u)&=\frac{\abs{\lambda_1}}{\pi}\ln\qty(\frac{1-u}{\abs{\lambda_1}}\frac{i\Im\lambda_1}{U(1-u,\lambda_1)+(1-u)\frac{\Re\lambda_1}{\abs{\lambda_1}}-\abs{\lambda_1}})\\
    &\qquad+\frac{\abs{\lambda_2}}{\pi}\ln\qty(\frac{u}{\abs{\lambda_2}}\frac{i\Im\lambda_2}{U(u,\lambda_2)+u\frac{\Re\lambda_2}{\abs{\lambda_2}}-\abs{\lambda_2}})\\
    &\qquad-\frac{1}{\pi}\ln\qty(\frac{i\Im\lambda_2}{U(u,\lambda_2)+u-\Re\lambda_2})
\end{align}
where
\begin{align}
    \abs{\lambda_1}\coloneqq\frac{\abs{k_-^A}+l}{\abs{k_-^C}+l}\qc\abs{\lambda_2}\coloneqq\frac{\abs{k_-^B}+l}{\abs{k_-^C}+l}\qc U(u,u')\coloneqq(u-u')\sqrt{\frac{u-\bar{u}'}{u-u'}}
\end{align}
and $\lambda_1,\lambda_2$ satisfy the relation
\begin{align}
    \lambda_1+\lambda_2=1.
\end{align}
Thus we can find the inverse local coordinate maps given by
\begin{align}
    (f_{(3,1)}^l)^{-1}(u)&=-\frac{1-u}{\abs{\lambda_1}}\frac{\Im\lambda_1}{U_-(1-u,\lambda_1)+(1-u)\frac{\Re\lambda_1}{\abs{\lambda_1}}-\abs{\lambda_1}}\\
    &\qquad\times\qty(\frac{u}{\abs{\lambda_2}}\frac{\Im\lambda_2}{U_+(u,\lambda_2)+u\frac{\Re\lambda_2}{\abs{\lambda_2}}-\abs{\lambda_2}})^{\frac{\abs{\lambda_2}}{\abs{\lambda_1}}}\\
    &\qquad\times\qty(\frac{\Im\lambda_2}{U_+(u,\lambda_2)+u-\Re\lambda_2})^{-\frac{1}{\abs{\lambda_1}}}\\
    (f_{(3,2)}^l)^{-1}(u)&=\qty(-\frac{1-u}{\abs{\lambda_1}}\frac{\Im\lambda_1}{U_+(1-u,\lambda_1)+(1-u)\frac{\Re\lambda_1}{\abs{\lambda_1}}-\abs{\lambda_1}})^{\frac{\abs{\lambda_1}}{\abs{\lambda_2}}}\\
    &\qquad\times-\frac{u}{\abs{\lambda_2}}\frac{\Im\lambda_2}{U_-(u,\lambda_2)+u\frac{\Re\lambda_2}{\abs{\lambda_2}}-\abs{\lambda_2}}\\
    &\qquad\times\qty(-\frac{\Im\lambda_2}{U_-(u,\lambda_2)+u-\Re\lambda_2})^{-\frac{1}{\abs{\lambda_2}}}\\
    (I\circ f_{(3,3)}^l)^{-1}(u)&=\qty(-\frac{u+1}{\abs{\lambda_1}}\frac{\Im\lambda_1}{uU_+(1+1/u,\lambda_1)+(u+1)\frac{\Re\lambda_1}{\abs{\lambda_1}}-\abs{\lambda_1}u})^{-\abs{\lambda_1}}\\
    &\qquad\times\qty(\frac{1}{\abs{\lambda_2}}\frac{\Im\lambda_2}{-uU_-(-1/u,\lambda_2)+\frac{\Re\lambda_2}{\abs{\lambda_2}}+\abs{\lambda_2}u})^{-\abs{\lambda_2}}\\
    &\qquad\times-u\cdot-\frac{\Im\lambda_2}{-uU_-(-1/u,\lambda_2)+1+\Re\lambda_2u}
\end{align}
and their derivatives are
\begin{align}
    \eval{\dv{(f_{(3,1)}^l)^{-1}}{u}}_{u=1}&=\frac{\Im\lambda_2}{2\abs{\lambda_1}^2}\qty(\frac{\Im\lambda_2}{\abs{\lambda_1}\abs{\lambda_2}+\Re\lambda_2-\abs{\lambda_2}^2})^{\frac{\abs{\lambda_2}}{\abs{\lambda_1}}}\qty(\frac{\Im\lambda_2}{\abs{\lambda_1}+\Re\lambda_1})^{-\frac{1}{\abs{\lambda_1}}}\label{derivative:Mandelstam1}\\
    \eval{\dv{(f_{(3,2)}^l)^{-1}}{u}}_{u=0}&=\qty(\frac{\Im\lambda_2}{\abs{\lambda_1}\abs{\lambda_2}+\Re\lambda_1-\abs{\lambda_1}^2})^{\frac{\abs{\lambda_1}}{\abs{\lambda_2}}}\frac{\Im\lambda_2}{2\abs{\lambda_2}^2}\qty(\frac{\Im\lambda_2}{\abs{\lambda_2}+\Re\lambda_2})^{-\frac{1}{\abs{\lambda_2}}}\label{derivative:Mandelstam2}\\
    \eval{\dv{(I\circ f_{(3,3)}^l)^{-1}}{u}}_{u=0}&=\qty(-\frac{\Im\lambda_1}{\abs{\lambda_1}+\Re\lambda_1})^{-\abs{\lambda_1}}\qty(\frac{\Im\lambda_2}{\abs{\lambda_2}+\Re\lambda_2})^{-\abs{\lambda_2}}\frac{\Im\lambda_2}{2}\label{derivative:Mandelstam3}
\end{align}
where
\begin{align}
    U_\pm(u,u')\coloneqq\pm\sqrt{(u-u')(u-\bar{u}')}.
\end{align}

Let us consider the limits $l\to\infty$ and $l\to0$. Since $\lambda_1,\lambda_2$ are given by
\begin{align}
    \lambda_1&=\frac{1}{2(1+L)^2}(L^2+2\tilde{\alpha}_A(2L+1))-\frac{i}{2(1+L)^2}\sqrt{L(2+3L)(2\tilde{\alpha}_A+L)(2\tilde{\alpha}_B+L)}\\
    \lambda_2&=\frac{1}{2(1+L)^2}(L^2+2\tilde{\alpha}_B(2L+1))+\frac{i}{2(1+L)^2}\sqrt{L(2+3L)(2\tilde{\alpha}_A+L)(2\tilde{\alpha}_B+L)}
\end{align}
where
\begin{align}
    L\coloneqq\frac{l}{\abs{k_-^C}},
\end{align}
the limit $l\to\infty$ corresponds to
\begin{align}
    \lambda_1&\to\frac{1}{2}-i\frac{\sqrt{3}}{2}\\
    \lambda_2&\to\frac{1}{2}+i\frac{\sqrt{3}}{2}.
\end{align}
After taking this limit, $\lambda_1$ and $\lambda_2$ do not depend on $k_-$ and this is the same as the Witten vertex. On the other hand, the limit $l\to0$ corresponds to
\begin{align}
    \lambda_1&\to\tilde{\alpha}_A\\
    \lambda_2&\to\tilde{\alpha}_B.
\end{align}
In this limit, equations \eqref{derivative:Mandelstam1},\eqref{derivative:Mandelstam2} and \eqref{derivative:Mandelstam3} become \eqref{derivative:Mandelstam} respectively. Therefore the Kaku vertex is intermediate between the Witten vertex and the light-cone vertex.

Similarly, the quartic Kaku vertex for the case $\mathrm{sgn}(k_-^A,k_-^B,k_-^C,k_-^D)=(+,-,+,-)$ or $(-,+,-,+)$, is defined by
\begin{align}
    V_4^l(A,B,C,D)&=\int_{m_-^l}^{m_+^l}\dd{m}\ev{b(v)f_{(4,1)}^l\circ A(0)f_{(4,2)}^l\circ B(0)f_{(4,3)}^l\circ C(0)f_{(4,4)}^l\circ D(0)}_\mathrm{UHP}\\
    &=(-1)^{1+\abs{C}}\omega(A,m_3^l(B,C,D)),
\end{align}
and, for the other cases, we define
\begin{align}
    V_4^l(A,B,C,D)=0.\label{def:quarticKaku}
\end{align}
As with the quartic light-cone vertex, the maps $f_{(4,r)}^l$ are also derived from Mandelstam mapping
\begin{align}
    \rho(u)=\int\dd{u}\frac{u-u_*}{u(u-m)(u-1)}
\end{align}
where the operators $A,B,C$ and $D$ are placed respectively at $1,m,0$ and $\infty$. The region $[m_-^l,m_+^l]$ of integration with respect to $m$ depends on $l$. It coincides with the quartic light-cone vertex region in the limit $l\to0$ and vanishes in the limit $l\to\infty$.

Therefore the action using the Kaku vertices is given by
\begin{align}
    S^l=\frac{1}{2}\omega(\Psi,Q\Psi)+\frac{1}{3}\omega(\Psi,m_2^l(\Psi,\Psi))+\frac{1}{4}\omega(\Psi,m_3^l(\Psi,\Psi,\Psi)).
\end{align}
We do not need to introduce higher order vertices because it has the $A_\infty$ structure and satisfies the BV master equation.

In this theory, it was shown that we do not face problems in the limit where all momenta become zero \cite{Erler:2020beb}. In the next section, when examining the cohomology around solutions, we need to consider the case where only one momentum becomes zero. For our purpose, we will confirm that there are no problems in any case in the following.

As in the previous section, let us consider
\begin{align}
    \lim_{k_-^A\to0}V_3^l(A,B,C)=\lim_{\tilde{\alpha}_A\to0}({f_{(3,1)}^l}'(1))^{h_A}({f_{(3,2)}^l}'(0))^{h_B}({I\circ f_{(3,3)}^l}'(0))^{h_C}\ev{A(1)B(0)C(\infty)}_\mathrm{UHP}.
\end{align}
The limit $\tilde{\alpha}_A\to0$ corresponds to
\begin{align}
    \begin{alignedat}{1}
        \lambda_1&\to\frac{1}{2(1+L)^2}L^2-\frac{i}{2(1+L)^2}L\sqrt{(2+3L)(2+L)},\\
        \lambda_2&\to\frac{1}{2(1+L)^2}(L^2+2(2L+1))+\frac{i}{2(1+L)^2}L\sqrt{(2+3L)(2+L)}.
    \end{alignedat}\label{limit:a0}
\end{align}
Hence we obtain
\begin{align}
    &\lim_{\tilde{\alpha}_A\to0}({f_{(3,1)}^l}'(1))^{h_A}({f_{(3,2)}^l}'(0))^{h_B}({I\circ f_{(3,3)}^l}'(0))^{h_C}\\
    &=\qty[\frac{\sqrt{(2+L)(2+3L)}}{4L}\qty(\frac{2+3L}{2+L})^{\frac{1+L}{L}}]^{-h_A}\\
    &\qquad\times\qty[\qty(\frac{\sqrt{(2+L)(2+3L)}}{2+L})^{\frac{L}{1+L}}\frac{(2+L)(2+3L)}{4(1+L)^2}]^{-h_B}\\
    &\qquad\times\qty[\qty(\frac{\sqrt{(2+L)(2+3L)}}{2+3L})^{-\frac{L}{1+L}}\frac{(2+L)(2+3L)}{4(1+L)^2}]^{-h_C}.\label{qubicKaku0}
\end{align}
This is finite and nonzero even for $h_A\neq0$. Further, if we consider $C$ to be $k_-^C=0$, the limit $k_-^C\to0$ corresponds to the limit $L\to\infty$ and equation \eqref{limit:a0} becomes
\begin{align}
    \begin{alignedat}{1}
        \lambda_1&\to\frac{1}{2}-i\frac{\sqrt{3}}{2}\\
        \lambda_2&\to\frac{1}{2}+i\frac{\sqrt{3}}{2}.
    \end{alignedat}\label{limit:ac0}
\end{align}
In this limit, the vertex is the same as the Witten vertex, and the result is unique, unlike the case in the previous section. In summary, if $A$ and $B$ are the states with $k_-^A=k_-^B=0$, the cubic Kaku vertex is the same as the Witten vertex.
\begin{align}
    m_2^l(A,B)=m_2^\mathrm{W}(A,B)\label{relation:KakuWitten}
\end{align}
Above properties hold true for descendant states as well. Therefore we can consider the states with $k_-=0$ without any problem in the Kaku theory.

A noteworthy feature is the limit $l\to0$. The Kaku vertex is, by definition, the same as the light-cone vertex in the limit $l\to0$. In fact, if we consider the limit $l\to0$ after we set $k_-^A=0$, it corresponds to the limit $L\to0$ and equation \eqref{limit:a0} becomes respectively
\begin{align}
    \begin{alignedat}{1}
        \lambda_1&\to0\\
        \lambda_2&\to1.
    \end{alignedat}\label{limit:al0}
\end{align}
This leads to the same result as the one in the previous section, and the cubic Kaku vertex with $l=0$ is not well-defined except for $h_A\ge0$. However if we take the limit $l\to0$ after we set $k_-^A=k_-^B=k_-^C=0$, the limit uniquely corresponds to \eqref{limit:ac0}. Thus equation \eqref{relation:KakuWitten} is valid for arbitrary $l$ including $l=0$.

\section{Classical solutions\label{cla}}
In the previous section, we have confirmed that the cubic Kaku vertex for the states with $k_-=0$ is the same as the cubic Witten vertex. In this section, we find classical solutions in the Kaku theory using this property.

The equation of motion in the Kaku theory is
\begin{align}
    0=Q\Psi+m_2^l(\Psi,\Psi)+m_3^l(\Psi,\Psi,\Psi).
\end{align}
If $\Psi$ is a state with $k_-=0$, because of \eqref{relation:KakuWitten} and \eqref{def:quarticKaku}, the equation of motion is
\begin{align}
    0=Q\Psi+m_2^l(\Psi,\Psi)+m_3^l(\Psi,\Psi,\Psi)=Q\Psi+m_2^\mathrm{W}(\Psi,\Psi).
\end{align}
Thus, any classical solutions with $k_-=0$ in the Witten theory satisfy the equation of motion in the Kaku theory.

Of course, even if we confirm that classical solutions in the Witten theory are also classical solutions in the Kaku theory, we do not know whether the physical interpretation and physical observables are the same. Here, we examine the action evaluated on the solution and the cohomology around the solution in the Kaku theory.

Using \eqref{relation:KakuWitten} and \eqref{def:quarticKaku}, the action in the Kaku theory evaluated on the solution $\Psi_\mathrm{sol}$ is
\begin{align}
    &\frac{1}{2}\omega(\Psi_\mathrm{sol},Q\Psi_\mathrm{sol})+\frac{1}{3}\omega(\Psi_\mathrm{sol},m_2^l(\Psi_\mathrm{sol},\Psi_\mathrm{sol}))+\frac{1}{4}\omega(\Psi_\mathrm{sol},m_3^l(\Psi_\mathrm{sol},\Psi_\mathrm{sol},\Psi_\mathrm{sol}))\\
    &=\frac{1}{2}\omega(\Psi_\mathrm{sol},Q\Psi_\mathrm{sol})+\frac{1}{3}\omega(\Psi_\mathrm{sol},m_2^\mathrm{W}(\Psi_\mathrm{sol},\Psi_\mathrm{sol})).\label{def:energy}
\end{align}
Since the action for a time independent solution is equal to minus the energy of the solution times the volume of the time coordinate, the energy of the time independent solutions with $k_-=0$ has the same value in either the Kaku theory or the Witten theory.

Next, let us consider the cohomology around classical solutions. The action around the classical solution in the Kaku theory is given by
\begin{align}
    S^{l,\mathrm{sol}}=\frac{1}{2}\omega(\Psi,Q^l_\mathrm{sol}(\Psi,\Psi))+\frac{1}{3}\omega(\Psi,m_2^{l,\mathrm{sol}}(\Psi,\Psi))+\frac{1}{4}\omega(\Psi,m_3^l(\Psi,\Psi,\Psi))
\end{align}
where
\begin{align}
    Q^l_\mathrm{sol}A&\coloneqq QA+m_2^l(\Psi_\mathrm{sol},A)+m_2^l(A,\Psi_\mathrm{sol})+m_3^l(\Psi_\mathrm{sol},\Psi_\mathrm{sol},A)+m_3^l(\Psi_\mathrm{sol},A,\Psi_\mathrm{sol})+m_3^l(A,\Psi_\mathrm{sol},\Psi_\mathrm{sol})\\
    m^{l,\mathrm{sol}}_2(A,B)&\coloneqq m_2^l(A,B)+m_3^l(\Psi_\mathrm{sol},A,B)+m_3^l(A,\Psi_\mathrm{sol},B)+m_3^l(A,B,\Psi_\mathrm{sol}).
\end{align}
Since $\Psi_\mathrm{sol}$ is a state with $k_-=0$ and $m_3^l$ is defined by \eqref{def:quarticKaku} for the states with $k_-=0$, we obtain
\begin{align}
    Q^l_\mathrm{sol}A&=QA+m_2^l(\Psi_\mathrm{sol},A)+m_2^l(A,\Psi_\mathrm{sol})\\
    m^{l,\mathrm{sol}}_2(A,B)&=m_2^l(A,B).
\end{align}
Thus, the action around the classical solution in the Kaku theory is 
\begin{align}
    S^{l,\mathrm{sol}}=\frac{1}{2}\omega(\Psi,Q\Psi+m_2^l(\Psi_\mathrm{sol},\Psi)+m_2^l(\Psi,\Psi_\mathrm{sol}))+\frac{1}{3}\omega(\Psi,m_2^l(\Psi,\Psi))+\frac{1}{4}\omega(\Psi,m_3^l(\Psi,\Psi,\Psi)).
\end{align}

In general, we do not know whether the cohomology of $Q^l_\mathrm{sol}$ is isomorphic to the cohomology of $Q^\mathrm{W}_\mathrm{sol}$. On the other hand,for the tachyon vacuum solution $\Psi_\mathrm{sol}=\Psi_\mathrm{tv}$, we can confirm that the cohomology of $Q^l_\mathrm{tv}$ is trivial. In the Witten theory, there exists a homotopy operator $A_\mathrm{W}$ which satisfies the following condition:
\begin{align}
    Q^\mathrm{W}_\mathrm{tv}A_\mathrm{W}=QA_\mathrm{W}+m_2^\mathrm{W}(\Psi_\mathrm{tv},A_\mathrm{W})+m_2^\mathrm{W}(A_\mathrm{W},\Psi_\mathrm{tv})=1.
\end{align}
Here we assume that the homotopy operator is the state with $k_-=0$. Usually, this assumption is satisfied \cite{Ellwood:2006ba,Erler:2009uj}. The existence of the homotopy operator proves that the cohomology of $Q^\mathrm{W}_\mathrm{tv}$ vanishes \cite{Ellwood:2006ba}. If the solution which we construct is a tachyon vacuum solution in the Kaku theory, it should have the homotopy operator $A_l$ which satisfies
\begin{align}
    Q^l_\mathrm{tv}A_l=1.
\end{align}
Using \eqref{relation:KakuWitten}, the homotopy operator in the Witten theory satisfies also
\begin{align}
    Q^l_\mathrm{tv}A_\mathrm{W}=QA_\mathrm{W}+m_2^\mathrm{W}(\Psi_\mathrm{tv},A_\mathrm{W})+m_2^\mathrm{W}(A_\mathrm{W},\Psi_\mathrm{tv})=1
\end{align}
Therefore, the tachyon vacuum solution $\Psi_\mathrm{tv}$ in the Witten theory satisfies also the equation of motion in the Kaku theory, and the cohomology of $Q^l_\mathrm{tv}$ is trivial.
\section{Summary\label{summary}}
In this paper, we have shown that the equation of motion with the restriction $k_-=0$ in the Kaku theory is equivalent to that in the Witten theory. Hence it allows us to use classical solutions with $k_-=0$ in the Witten theory as classical solutions in the Kaku theory. The energy of the solution has the same value in either the Kaku theory or the Witten theory, and for the tachyon vacuum solution, there exists the homotopy operator. However, we have not given a complete proof that the physical interpretation is the same.

We have considered the Kaku theory but have not set the Chan-Paton parameter $l$ to a specific choice. Thus, one may expect that we can construct the tachyon vacuum solution and show the existence of the homotopy operator even in the theory using the light-cone vertex $l\to0$. In fact, because equation \eqref{def:energy} holds even in the limit $l\to0$, the energy of the time independent solutions with $k_-=0$ is the same in either the theory using the light-cone vertex or the Witten theory. However, in the limit $l\to0$, the shifted kinetic operator $Q^l_\mathrm{sol}$ is not well-defined since the cubic Kaku vertex with $l\to0$ is not well-defined due to \eqref{limit:al0}. Hence, $Q^{l\to0,\mathrm{sol}},m_2^{l\to0,\mathrm{sol}}$ and $S^{l\to0,\mathrm{sol}}$ are not well-defined, and we cannot consider the cohomology around the classical solution.

We consider the state with $k_-=0$ in the Kaku theory and obtain some properties. In particular, the properties for the state with $k_-=0$
\begin{alignat}{2}
    m_2^l(A,B)&=m_2^\mathrm{W}(A,B)&&\qfor k_-^A=k_-^B=0\\
    m_3^l(A,B,C)&=0&&\qfor k_-^A=k_-^B=k_-^C=0
\end{alignat}
play an important role. These properties allow us to consider the Witten vertex instead of the Kaku vertex in constructing classical solutions and examining the energy of them. Using these properties, we may understand more in the Kaku theory besides what is considered in this paper by analogy with the Witten theory.

We confirmed the energy of the classical solution with $k_-=0$ but we have not examined all the physical observables. The observable
\begin{align}
    \omega(\Psi_\mathrm{sol},m_0^\mathrm{W})
\end{align}
exists in the Witten theory where $m_0^\mathrm{W}$ is an on-shell closed string state, and this is called Ellwood invariant \cite{Gaiotto:2001ji,Ellwood:2008jh}. To define the Ellwood invariant in the Kaku theory, we need to construct $m_0^l$ which satisfies 
\begin{align}
    0&=m_2^l(m_0^l,A)+(-1)^\abs{A}m_2^l(A,m_0),\\
    0&=m_3^l(m_0^l,A,B)+(-1)^\abs{A}m_3^l(A,m_0,B)+(-1)^{\abs{A}+\abs{B}}m_3^l(A,B,m_0).
\end{align}
It is an open problem whether such $m_0^l$ exists and the gauge invariant which is defined using $m_2^l,m_3^l$ and $m_0^l$ is equivalent to the Ellwood invariant. We expect that solving the problem is unavoidable to evaluate closed string amplitudes because the Ellwood invariant in the Witten theory is a closed string tadpole.
\section*{Acknowledgements}
The author would like to thank Nobuyuki Ishibashi for reading a preliminary draft and helpful comments.
This work was supported by JST, the establishment of university fellowships towards the creation of science technology innovation, Grant Number JPMJFS2106.
\printbibliography
\end{document}